\documentclass{osa-article}

\journal{oe} 

\articletype{Research Article}

\usepackage{lineno}
\usepackage[T1]{fontenc}
\usepackage{tabularx}
\usepackage{blkarray}
\usepackage{multirow}
\usepackage{braket}

\def\blue#1{{\color{black} #1 \color{black}}}

\begin{document}

\title{Experimental hierarchy of the nonclassicality of single-qubit states via potentials \blue{for} entanglement, steering, and Bell nonlocality}

\author{Josef Kadlec,\authormark{1} Karol Bartkiewicz,\authormark{2} Antonín \v{C}ernoch,\authormark{3} Karel Lemr,\authormark{1,*} and Adam Miranowicz\authormark{2}}

\address{\authormark{1}Palack\'{y} University in Olomouc, Faculty of Science, Joint Laboratory of Optics of Palack\'{y} University and Institute of Physics AS CR, 17. listopadu 12, 771 46 Olomouc, Czech Republic\\
\authormark{2}Institute of Spintronics and Quantum Information, Faculty of Physics, Adam Mickiewicz University, 61-614 Pozna\'{n}, Poland\\
\authormark{3}Institute of Physics of the Academy of Sciences of the Czech Republic, Joint Laboratory of Optics of Palack\'{y} University and Institute of Physics AS CR, 17. listopadu 50a, 772 07 Olomouc, Czech Republic}

\email{\authormark{*}k.lemr@upol.cz} 


\begin{abstract*} 
Entanglement potentials are a promising way to quantify the nonclassicality of single-mode states. They are defined by the amount of entanglement (expressed by, e.g., the Wootters concurrence) obtained after mixing the examined single-mode state with a purely classical state; such as the vacuum or a coherent state. We generalize the idea of entanglement potentials to other quantum correlations: the EPR steering and Bell nonlocality, thus enabling us to study mutual hierarchies of these non-classicality potentials. Instead of the usual vacuum and one-photon superposition states, we experimentally test this concept using specially tailored polarization-encoded single-photon states. One polarization encodes a given non-classical single-mode state, while the other serves as the vacuum place-holder. This technique proves to be experimentally more convenient in comparison to the vacuum and a one-photon superposition as it does not require the vacuum detection. 

\end{abstract*}

\section{Introduction}

Nonclassical quantum states play a crucial role in many modern quantum technologies. 
In particular, single-photon states allow to create a secure communication channel \cite{Bennett_2014,Shor2000}. Entangled photon pairs even permit to develop a protocol for secure communication using untrusted devices \cite{Ekert1991}.    
In quantum metrology, one can achieve super-resolution \cite{Jacobson1995,Fonseca1999}, i.e., resolution beyond the wavelength  limit, using, for example, highly non-classical NOON states; and recently, squeezed states of light have been used to enhance detection of gravitational waves 
\cite{Grote2013,Aasi2013,Tse2019,Acernese2019}
 and to show quantum advantage by implementing boson sampling with squeezed states \cite{Zhong2020,Madsen2022}.

Given these applications, quantifying the nonclassicality of light is of paramount importance.
Over the years, many approaches to classify and quantify nonclassicality have been proposed. In quantum-optical systems, the non-positivity of the Glauber-Sudarshan $P$ function serves as a nonclassicality universal criterion \cite{glauberbook, vogelbook}. To quantify nonclassicality, various measures were introduced including the nonclasical depth \cite{Hillery1987} or the nonclassical distance \cite{Lee1991,	Luetkenhaus1995}.

\begin{figure}[h!]
\centering
\includegraphics[scale=0.8]{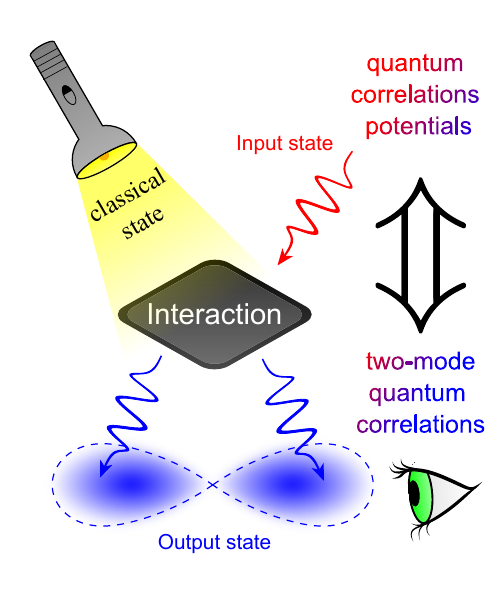}
\caption{The conceptual scheme for the quantification of single-mode nonclassicality via the observation of two-mode quantum correlations.}
\label{fig_conc_scheme}
\end{figure}

In this paper, we use a conceptually different approach towards nonclassicality (NC) quantification based on the so-called NC potentials \cite{Miranowicz2023}.
The NC potential of a single-mode quantum state is given by the amout of NC correlations (e.g. entanglement) that it is capable of producing upon interacting with a classical state (e.g. the vacuum or a coherent state) as illustrated by a conceptual scheme in Fig.~\ref{fig_conc_scheme}. A straightforward example of this concept is a single-photon state mixed on a balanced beam splitter with the vacuum state producing a maximally entangled two-qubit state.
This method benefits from a well-established theory for quantifying quantum correlations of bipartite systems to characterize NC of a single-mode system.
We start from the entanglement potential of Asb\'{o}th \emph{et al.} \cite{Asboth2005}, extending it, following our theoretical work
  \cite{Miranowicz2023}, with more NC correlations, namely by EPR steering \cite{Cavalcanti_2016} and Bell nonlocality \cite{Brunner2014}.
   This allows us to establish the hierarchy of single-qubit
    nonclassicalities using the well-known two-qubit NC correlations
     hierarchy \cite{Jones2007}.
\blue{Note that hierarchies of quantum correlations have been recently analysed in conceptually different quantum systems.
These include the hierarchies of spatial correlations of two-qubit states \cite{Jirakova2021,Abo2023, Fan2023} and two-mode Gaussian states \cite{Qureshi2018}, and the hierarchies of temporal correlations \cite{Ku2018,Ku2022,Ullah19}.}

We have constructed an experimental setup on the platform of linear optics, because photons are a suitable test bed. 
The most typical classical single-mode state would be the vacuum and hence the theoretical framework is usually developed for the so-called vacuum and one-photon superposition (VOPS) states, including both pure and mixed states. The vacuum detection, however, proves experimentally challenging and requires homodyne detection \cite{Jakeman1975}.
To avoid this experimental intricacy, we have implemented an alternative method using a 
polarization encoding of photon pairs. Single-photon states are encoded into the horizontal-polarization mode, while the vertical-polarization mode serves as a place-holder for the vacuum (corresponding to no photon in the horizontal-polarization mode).

 To compensate for physically different behavior between vertically polarized photons and the real vacuum state, we have developed a detection setup to reconstruct 
density matrices by a two-photon coincidence detection. The polarization encoding together with this detection procedure allows us to observe all relevant NC effects in complete analogy to the genuine VOPS states. For detailed description see Sec.~\ref{sec_setup}.

\begin{figure}[!t]
\centering
\includegraphics[width=\textwidth]{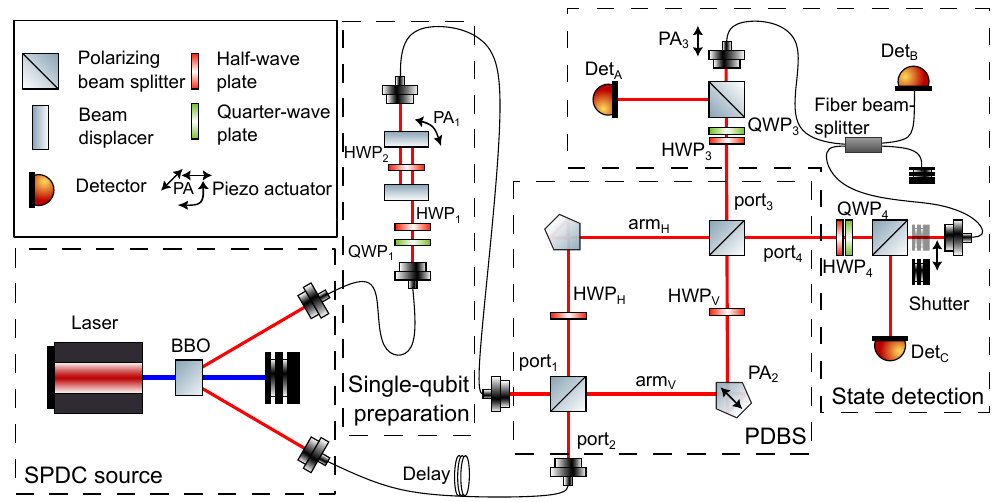}
\caption{Scheme of the experimental setup as explained in the text. SPDC denotes spontainous parametric down conversion, BBO is a beta barium borate ($\beta$-BaB\textsubscript{2}O\textsubscript{4}) crystal and PDBS stands for a polarization dependent beam splitter.}
\label{setup}
\end{figure}

\section{Nonclassicality potentials}\label{sec_NP_pot}

In this paper, we calculate NC potentials of experimentally generated states from previously estimated output two-mode density matrices. Estimation of density matrices uses a maximum likelihood method as explained in Appendix B.

In the first step a single-qubit VOPS state in the form of
\begin{equation}\label{sigmapx}
\sigma(p,y) = \left(
\begin{array}{cc}
1-p & y\\
y^* &p
\end{array}\right),
\end{equation}
is generated, expressed in the basis of the vacuum $\ket{\text{vac}}$ and  the Fock single-photon state $\ket{1}$. Thus, $p$ denotes a single-photon probability, and $y$ is a coherence parameter.

\subsection{Perfect beam splitter}

The examined state is then mixed with a pure vacuum state on a balanced lossless beam splitter (BS), resulting in a two-qubit state,
\begin{equation}\label{rhopx}
\rho(p,y) = \left(
\begin{array}{cccc}
1-p & -\frac{1}{\sqrt{2}}y & \frac{1}{\sqrt{2}}y& 0\\
 -\frac{1}{\sqrt{2}}y^* & \frac{1}{2}p &  -\frac{1}{2}p &0\\
 \frac{1}{\sqrt{2}}y^* & -\frac{1}{2}p &  \frac{1}{2}p &0\\
0&0&0&0 
\end{array}
\right) .
\end{equation}
From this matrix one may calculate any type of two-qubit quantum correlation. The state $\sigma(p,y)$ is then attributed NC potentials based on the corresponding quantum correlations observed in $ \rho(p,y)$. In the next sections, we focus on the measures of entanglement, EPR steering, and Bell's nonlocality.

As an entanglement measure, we use the Wootters concurrence \cite{Wootters1998}: by defining a matrix $\tilde{\rho}= \rho(\sigma_2 \otimes \sigma_2)\rho^*(\sigma_2 \otimes \sigma_2)$, where $\sigma_2$ is the Pauli Y-matrix, and its eigenvalues (sorted in ascending order), $\lambda_i^2 = \text{eig}(\tilde{\rho})$, the concurrence takes form
$C(\rho)= \text{max}\big(0, \lambda_1-\lambda_2-\lambda_3-\lambda_4\big)$.

To quantify steering, we use the Costa-Angelo steering measure for a three-measurement scenario \cite{Costa2016}. Using the Bloch representation, a general two-qubit state can be expresed as

\begin{equation}
\rho = \frac{1}{4}\big( I\otimes I + 
\boldsymbol{u}\cdot\boldsymbol{\sigma}\otimes I 
+ I\otimes \boldsymbol{v}\cdot \boldsymbol{\sigma}
+ \sum\limits_{n,m=1}^{3} T_{mn}\sigma_n\otimes\sigma_m  \big),
\end{equation}
where $\boldsymbol{\sigma} = [\sigma_1,\sigma_2,\sigma_3]$ is a vector of the Pauli matrices and $I$ is $2\times 2$ identity matrix. 
Then the steering measure takes the form
\begin{equation}
S(\rho) = \text{max}\Big(0,\frac{\sqrt{\text{Tr}R}-1}{\sqrt{3}-1}\Big),
\end{equation}
where $T = [T_{mn}]$ and $R= T^TT$ are correlation matrices.

To measure the nonlocality we use a similar formula. Bell-nonlocality can be regarded as a Costa-Angelo steering in a two-measure scenario \cite{Costa2016}, so the value of the corresponding Bell nonlocality measure can be given by formula
\begin{equation}
B(\rho) = \text{max}\Big(0,\frac{\sqrt{\text{Tr}R-\text{min}[\text{eig}(R)]}-1}{\sqrt{2}-1}\Big).
\end{equation}

Thus, we define NC potentials \cite{Miranowicz2023} as
\begin{align}
\text{CP}[\sigma(p,y)] &= \text{C}[\rho(p,y)],\\
\text{SP}[\sigma(p,y)] &= \text{S}[\rho(p,y)],\\
\text{BP}[\sigma(p,y)] &= \text{B}[\rho(p,y)],
\end{align}
where CP, SP and BP are the concurrence, steering, and Bell nonlocality potentials, respectively.

\subsection{Imperfect beam splitter}

In typical experimental implementations, due to unavoidable imperfections, the BS cannot be perfectly balanced and the output modes perfectly coherent. Let $r$ and $t$ be the BS reflection and transmission coefficients ($r^2+t^2=1$), respectively. While $r$ and $t$ are in general complex, their phases do not affect the calculated quantities, hence we assume them to be real and non-negative. Parameter $w$ characterizes the output decoherence (where $w = 1$ means a perfect coherence). Then the single-qubit state $\sigma(p,y)$ mixed with the vacuum on such an imperfect BS yields the two-qubit state \cite{Miranowicz2023}
\begin{equation}\label{rhorqpx}
\rho_{wr}(p,y) = \left(
\begin{blockarray}{cccc}
1-p & -wry & wty& 0\\
 -\frac{1}{\sqrt{2}}y^* & pr^2 &  -pw^2rt &0\\
 \frac{1}{\sqrt{2}}y^* & -pw^2rt &  pt^2 &0\\
0&0&0&0 
\end{blockarray}
\right).
\end{equation}
In the case of the perfectly balanced ($r=t=1/\sqrt{2}$) and non-decohering ($w=1$) BS, this equation reduces to Eq.~(\ref{rhopx}).

Similarly to the ideal case, we define NC potentials for the imperfect BS \cite{Miranowicz2023} as
\begin{align}
\text{CP}_{wr}[\sigma(p,y)] &= \text{C}[\rho_{wr}(p,y)],\\
\text{SP}_{wr}[\sigma(p,y)] &= \text{S}[\rho_{wr}(p,y)],\\
\text{BP}_{wr}[\sigma(p,y)] &= \text{B}[\rho_{wr}(p,y)].
\end{align}

\section{Experimental setup}\label{sec_setup}

\begingroup

\renewcommand{\arraystretch}{1.3} 

\begin{table*} \caption{\label{experimental_table}Summary of the experimental results.} 
\centering
\begin{tabularx}{\textwidth}{>{\raggedright\arraybackslash}X>{\raggedright\arraybackslash}X>{\raggedright\arraybackslash}Xl>{\raggedright\arraybackslash}X>{\raggedright\arraybackslash}X>{\raggedright\arraybackslash}X}
\hline \hline
State & $\text{Concurrence}$ & $\text{Steering}$ & $\text{Bell nonlocality}$ & $\text{Fidelity }\blue{F_{\text{in}}}$ & $\text{Fidelity }\blue{F_{\text{out}}}$\\ [-.5ex]
\hline
$\rho_1$ & $0.013^{+0.001}_{-0.001}$ & $0^{+0}_{-0} $& $0^{+0}_{-0}$ &$ 0.954^{+0.005}_{-0.005}$ & $0.954^{+0.005}_{-0.005}$\\ 
\hline
$\rho_2$ & $0.19^{+0.01}_{-0.01}$ &$0.038^{+0.001}_{-0.038}$ & $0.027^{+0.001}_{-0.027}$ &$ 0.970^{+0.005}_{-0.018}$ &$ 0.955^{+0.005}_{-0.016}$\\
\hline
$\rho_3$ & $0.43^{+0.02}_{-0.02}$ & $0.07^{+0.09}_{-0.07}$ &$ 0^{+0.11}_{-0} $& $0.91^{+0.04}_{-0.06} $& $0.89^{+0.04}_{-0.07}$\\
\hline
$\rho_4$ & $0.47^{+0.02}_{-0.02}$ & $0^{+0}_{-0}$ & $0^{+0}_{-0} $&$ 0.995^{+0.003}_{-0.018}$ &$ 0.965^{+0.007}_{-0.013}$\\
\hline
$\rho_5$ & $0.48^{+0.01}_{-0.01}$ & $0.14^{+0.04}_{-0.05}$ &$ 0.04^{+0.05}_{-0.04}$ & $0.93^{+0.02}_{-0.03}$ & $0.89^{+0.02}_{-0.02}$\\
\hline
$\rho_6$ &$ 0.52^{+0.01}_{-0.01} $& $0.22^{+0.02}_{-0.09} $& $0.17^{+0.02}_{-0.11} $&$ 0.93^{+0.01}_{-0.05} $& $0.90^{+0.01}_{-0.05}$\\
\hline
$\rho_7$ & $0.58^{+0.01}_{-0.01} $& $0.34^{+0.03}_{-0.03}$ &$ 0.26^{+0.05}_{-0.03}$ &$ 0.95^{+0.01}_{-0.01} $& $0.92^{+0.01}_{-0.01}$\\
\hline
$\rho_8$ & $0.68^{+0.02}_{-0.01}$ & $0.26^{+0.07}_{-0.06}$ &$ 0.11^{+0.09}_{-0.06}$ & $0.992^{+0.003}_{-0.010} $& $0.90^{+0.03}_{-0.03}$\\
\hline
$\rho_9$ & $0.70^{+0.01}_{-0.01} $& $0.53^{+0.02}_{-0.03}$ &$ 0.47^{+0.04}_{-0.06}$ &$ 0.966^{+0.005}_{-0.015}$ &$ 0.920^{+0.008}_{-0.014}$\\
\hline
$\rho_{10}$ & $0.85^{+0.01}_{-0.01}$ &$ 0.74^{+0.04}_{-0.05}$ &$ 0.68^{+0.07}_{-0.05}$ & $0.996^{+0.002}_{-0.057} $& $0.94^{+0.02}_{-0.04}$\\
\hline \hline
\end{tabularx}
\end{table*}

\endgroup

\begin{figure*}
\begin{center}
{\includegraphics[width=0.8\textwidth]{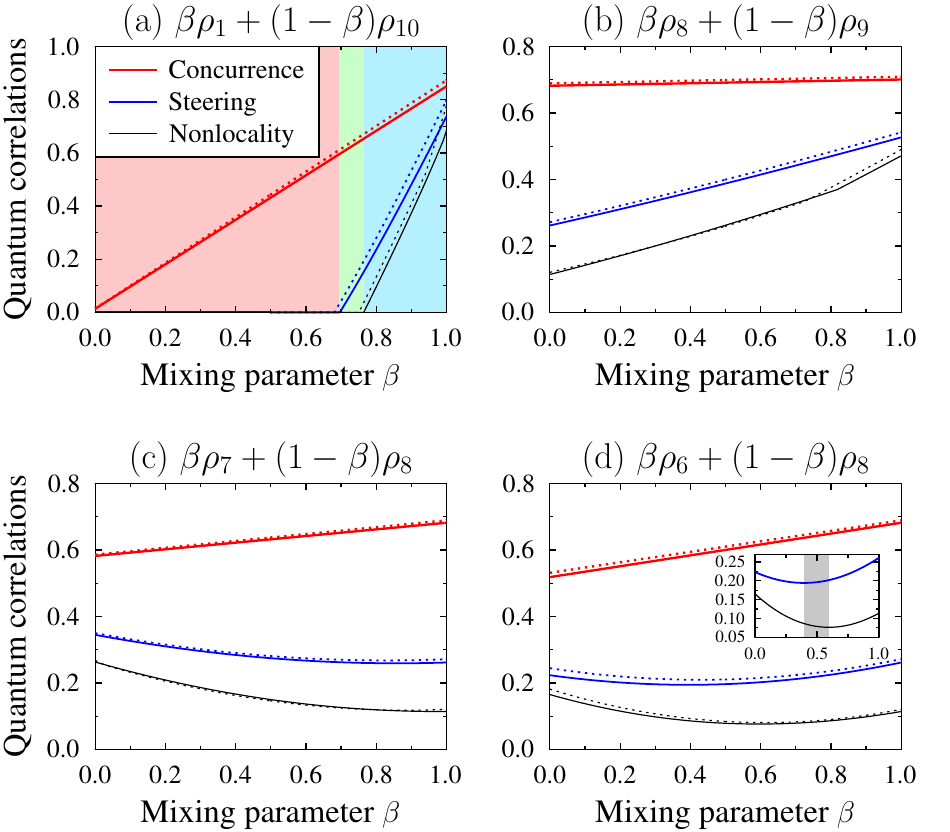}}\hspace*{0pt} 
\caption{Hierarchy of the quantum correlations described by the
measures of entanglement, $C[\rho(\beta)]$, EPR steering, $S[\rho(\beta)]$, and Bell nonlocality, $B[\rho(\beta)]$, for the states
$\rho(\beta)=\beta\rho_i+(1-\beta)\rho_j$, generated by numerically interpolating between chosen
pairs of experimental states $\rho_i$ and $\rho_j$. Solid curves depict interpolations between the experimental states while dotted curves correspond to the interpolations between the fitted states. In all cases the curves are ordered from top to bottom: the concurrence, steering, and Bell nonlocality potentials.}
 \label{mixing2}
 \end{center}
 \end{figure*}

\begin{figure*}[ht]
\begin{center}
{\includegraphics[width=1\textwidth]{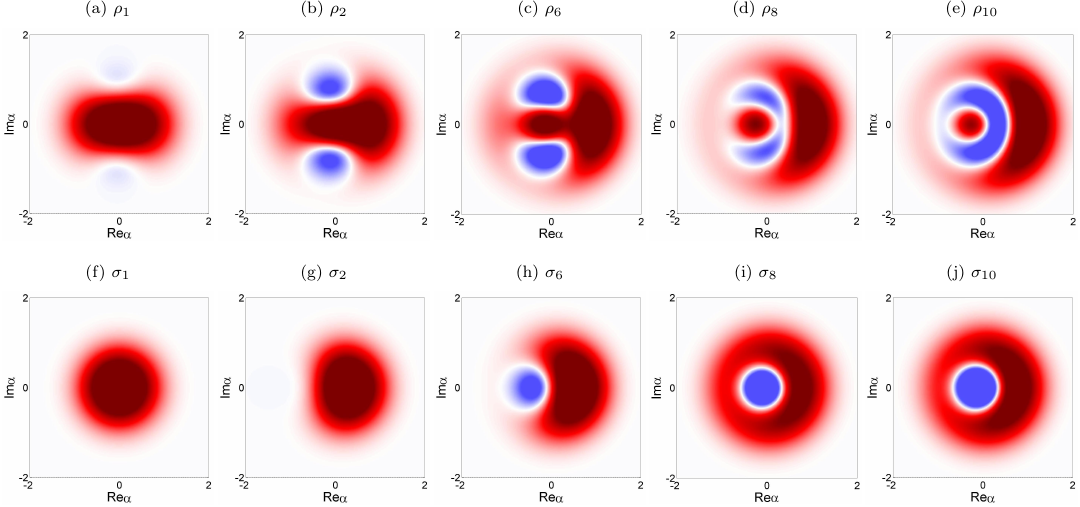}}\hspace*{0pt}\\ %
\caption{Wigner functions: $W^{(2)}(\alpha)$ for chosen
experimental two-qubit states $\rho_i$ ($i=1,2,6,8,10$) being
treated formally as qutrit states (in row 1), and
$W^{(1)}(\alpha)$ for the corresponding single-qubit states
$\sigma_i$ (in row 2). The theoretical qubit states $\sigma_i$
result in the qutrit states $\rho_i'$, which are the closest to
$\rho_i$. The darker red, the larger positive values of the Wigner
functions; while the darker blue, the larger their negative
values. The negative regions (marked by blue) of the Wigner
functions show clearly the nonclassicality of the represented
states. Note that $\sigma_1$ and $\sigma_2$ are nonclassical,
although their Wigner function appear nonnegative (considering margins of experimental noise).}
 \label{wigner}
 \end{center}
 \end{figure*}

The experimental setup is constructed on the platform of linear optics (see Fig.~\ref{setup}). Pairs of photons in a separable state are generated using type-I spontaneous parametric down conversion (SPDC).
All measured results are post-selected on the coincidence detection of both generated photons.
As mentioned in the Introduction, our experimental implementation avoids the problem of the vacuum detection. 
Instead of encoding qubits into the vacuum ($\ket{\text{vac}}$) and Fock ($\ket{1}$) states, we use vertical ($\ket{V}$) and horizontal ($\ket{H}$) polarization of photons, respectively.
One can think of the vertical polarization as a sort of place-holder or a witness for the vacuum in the horizontal mode.
Obviously, the vacuum and the vertically polarized photon behave quite differently on a beam splitter. As explained below, our detection procedure allows us to mitigate these differences.

The first photon in a generated pair is guided
 into the ``single-qubit preparation'' stage via an optical fiber (all optical fibers are single mode and most of them are equipped with polarization controllers, that are not depicted in the scheme). 
There, any pure qubit state is encoded into the photon polarization state using a set of motorized half-wave and quarter-wave plates. Subsequently, the pure qubit state is dephased to any degree using two beam displacers with a half-wave plate (HWP\textsubscript{2}) in between  rotated to 45° (all wave-plate rotations are expressed in relation to the horizontal-polarization direction).
 One beam displacer is mounted on a piezo actuator. 
Voltage is applied to this piezo actuator (PA\textsubscript{1}) in such a way that either the identity operation or a phase flip is imposed to the qubit state. 
The degree of dephasing is proportional to the probability ratio between these two actions.
This produces a qubit state in the form of Eq.~(\ref{sigmapx}).

Next, the qubit is guided via another fiber into one input port of a tunable polarization-dependent beam splitter (PDBS). The second photon in a generated pair is guided directly from the SPDC source using an optical fiber into the second input port of the PDBS. The second photon is fixed to the vertical polarization to play the role of the vacuum, considered here as a purely classical state.

The PDBS is implemented by means of a Mach-Zehnder interferometer consisting of two polarizing beam splitters with a motorized half-wave plate (HWP\textsubscript{H,V}) in each arm.
The horizontally (vertically) polarized components of both photons meet in the interferometer arm labeled arm\textsubscript{H} (arm\textsubscript{V}), considering a bit flip operation on the first photon by HWP\textsubscript{2} rotated by 45°. The splitting ratio $R_{\text{H;V}} = \blue{\sin}^2(2\theta_{\text{H;V}}) $ for the H and V modes (polarizations) is tuned by the motorized half-wave plates HWP\textsubscript{H} and HWP\textsubscript{V} rotated by appropriate angles $\theta_{\text{H;V}}$. To reverse the initial bit-flip operation on the first photon, another bit-flip operation is implemented in the output port\textsubscript{3} simply by biasing the rotation of HWP\textsubscript{3} by 45°.
Note that HWP\textsubscript{3} is used for both aforementioned transformation and also in our state analysis.

A specially designed detection apparatus is used for our output-state analysis. 
It consists of a motorized half and quarter-wave plates and a polarizing beam splitter in each output port of the interferometer. A fiber beam splitter, a shutter, and a piezo actuator are connected as depicted, to measure coherence terms across the output arms.  Three detectors are placed at the end of the setup. Events corresponding to two of these detectors clicking simultaneously are registered using a coincidence logic. This detection apparatus allows us to reconstruct the entire output density matrix $\rho_{\text{out}}$, as prescribed in Eq.~(\ref{rho_out}).
Due to unavoidable experimental imperfections, $\rho_{\text{out}}$ does not have the form of $\rho_{wr}(p,y)$ from Eq.~(\ref{rhorqpx}), nevertheless the closest state of such form has the Bures distance \cite{Vendral1997} typically lower than 0.01.
See Appendix~A for a more detailed description of the measurement procedure.

\section{Experimental results}\label{sec_results}

We used the experimental setup to generate a number of different single-photon states. Subsequently we applied the procedure mentioned in Appendix B to estimate for each state its density matrix $\rho_i$ ($i=1,2\ldots 10$), from which we computed the three nonclassicality measures: (a) the concurrence potential (b) the steering potential in a three-measurement scenario and (c) the Bell nonlocality potential. All of their values are given in Table~\ref{experimental_table}. 
To verify that the setup operates as expected, two fidelities \blue{($F(\rho_A,\rho_B) = [\text{Tr}\sqrt{\sqrt{\rho_A}\rho_B\sqrt{\rho_A}}]^2$)} were calculated for every generated state. The fidelity labelled F$_{\text{out}}$ \blue{= $F[\rho_i,\rho_\text{th}]$}describes the overlap between the estimated output state density matrix \blue{$\rho_i$}and the theoretical density matrix \blue{$\rho_\text{th} = \rho_{wr}(p,y)$ with parameters $\{p,y,w,r\}$}assuming a perfect setup operation.
Subsequently, minimizing the Bures distance measure, we fitted the closest states in the form of Eq.~(\ref{rhorqpx}), to the experimental $\rho_i$. These fitted states are defined by the four parameters \{$\bar{p},\bar{y},\bar{r},\bar{w}$\}, where $\bar{p}$ and $\bar{y}$ parametrize the single-qubit state defined in Eq.~(\ref{sigmapx}), and $\bar{r}$ and $\bar{w}$ characterise the beam splitter reflectivity and output decoherence, respectively. The fidelity F$_{\text{in}}$ = \blue{$F[\sigma(\bar{p},\bar{y}),\sigma(p,y)]$}quantifies the overlap between the fitted single-qubit \blue{$\sigma(\bar{p},\bar{y})$}input states and the expected input state assuming a perfect single-qubit preparation \blue{$\sigma(p,y)$.}
More properties of the experimental states have been calculated as we show in the Dataset 1 (Ref. \cite{DigitalSupplement}).

As expected, for all the generated states $\rho_i$, the relation $C(\rho_i)\geq S(\rho_i)\geq B(\rho_i)$ holds. 
One may notice, from looking at the concurrence column, that all of the two-qubit states are entangled. One can easily deduce from Eq.~(\ref{rhorqpx}) and the positive partial transpose condition, that a separable output state is in the theory only achieved by having the vacuum (vertically polarized photon) as the input state or by having maximally unbalanced or maximally decohering BS ($w=0 \lor r=0 \lor t=0$). 
Due to unavoidable experimental imperfections, including detection noise, this is experimentally unattainable.
In addition to the vacuum, which is the only separable state (for which $C=S=B=0$), the generated states belong to all physically valid categories, i.e., entangled non-steerable ($C>0; S=B=0$); steerable Bell local ($C\geq S>0=B$) and Bell nonlocal ($C\geq S \geq B>0$).

When comparing $C$, $S$, and $B$ of two different states, some counter-intuitive relations can be observed.
The generated states allow to observe the complexity of the hierarchy of quantum correlations.
For example, a state of higher $C$ (e.g. $\rho_4$) might exhibit lower degrees of $S$ and $B$ in comparison to another state (e.g. $\rho_2$).

These intricacies become even more pronounced when depicting correlations of interpolated mixtures of pairs of generated states. More specifically, the interpolations of pairs of states are depicted in Fig.~\ref{mixing2}. Figure~\ref{mixing2}(a) shows a typical transition of NC correlations between the least and most entangled experimentally obtained states ($\rho_1$ and $\rho_{10}$). This plot allows us to observe the expected hierarchy of the $C$, $S$ and $B$ correlations, specifically  the red area corresponds to entangled, but nonsteerable and local states; the green area encompasses entangled, steerable, and Bell local states; the blue area involves the Bell nonlocal states. 
In Fig.~\ref{mixing2}(b), the situation of steeply growing $S$ and $B$ across the interpolation range between $\rho_8$ and $\rho_9$ is depicted while, $C$ remains almost constant.
In contrast to that, the interpolation between $\rho_7$ and $\rho_8$, depicted in Fig.~\ref{mixing2}(c), exhibits a growing $C$ and simultaneously decreasing both $S$ and $B$.

While one can readily expect quantum correlations to be non-monotonic across the interpolation range with a minimum in the middle of the interval, we demonstrate a different position for these minima in the case of $S$ and $B$ as depicted in Fig.~\ref{mixing2}(d).
As a direct consequence, there is an interpolation range, highlighted by the gray area in the inset, where $B$ decreases while $S$ increases.

\blue{
Figure 4 shows the Wigner functions calculated using the formula \cite{gerry2005}
\begin{equation}
W(\alpha) = 
\frac{1}{\pi^2}\int \text{d}^2\lambda 
e^{\alpha\lambda^*-\alpha^*\lambda} C_W(\lambda),
\end{equation}
where $C_W(\lambda) = \text{Tr}[\rho e^{\lambda a^\dag-\lambda^*a}]$ is the characteristic function (not to be confused with concurrence $C[\rho]$), while $a$ and $a^\dag$ are the photon annihilation and creation operator, respectively.}Figure \ref{wigner} in row 1 shows the Wigner functions for the
chosen experimentally reconstructed two-qubit states $\rho_i$,
which are formally represented as qutrit states with the standard
encoding $\ket{00}\rightarrow \ket{0}$, $\ket{01}\rightarrow
\ket{1}$, and $\ket{10}\rightarrow \ket{2}$. \blue{We use this formally single-mode representation of two-mode states
for its graphical compactness. Note that standard visualization of
a two-mode Wigner function, say $W(\alpha_1,\alpha_2)$,
corresponds to showing its four marginal functions: $W({\rm
Re}\,\alpha_1,{\rm Re}\,\alpha_2)$, $W({\rm Re}\,\alpha_1,{\rm
Im}\,\alpha_2)$, $W({\rm Im}\,\alpha_1,{\rm Re}\,\alpha_2)$, and
$W({\rm Im}\,\alpha_1,{\rm Im}\alpha_2)$, as they are plotted in
our related paper~\cite{Miranowicz2023}.}Row 2 of Fig.
\ref{wigner} shows the Wigner functions for the corresponding single-qubit states $\sigma_i(p,y)$, for which the corresponding state
$\rho^{(i)}_{wr}(p,y)$, given in Eq.~(\ref{rhorqpx}), minimizes
the Bures distance to $\rho_i$ for numerically optimized values of
the parameters $w,r,p$, and $|y|\in[0,1]$.
The drawback of the Wigner function is that it does not universally detect the nonlocality of a state, as can be seen from the Wigner functions of $\sigma_1$ and $\sigma_2$. On the other hand, the method of nonlocality potentials proves to be more versatile, because it not only correctly identified both of these states as nonlocal, but also allows a more profound insight into the nonlocality due to the hierarchies: CP($\sigma_1$)>0, SP($\sigma_1)$=BP($\sigma_1$)=0 while CP($\sigma_2$)>SP($\sigma_2$)>BP($\sigma_2$)>0. 


\section{Conclusions} \label{sec_conc}

We have proposed and experimentally implemented an efficient
method, which was used to simulate the interaction of the vacuum
with VOPS states on a balanced beam splitter using only
polarization qubits. 
\blue{This implementation replaces the traditional vacuum and single photon encoding by polarization-state encoding. As a result, it can be employed more easily and even in near-future practical quantum networks, where a high-fidelity detection of the vacuum is technically very challenging.
Moreover, it might simplify the past quite demanding experiments with the VOPS states, which include quantum teleportation \cite{Lombardi2002} or EPR steering \cite{Fuwa2015}.}\blue{Our method constitutes an experimentally friendly framework  for various further research applications of single-mode state nonclassicality.}Compared to the setups applied by us in
Refs.~\cite{Jirakova2021,Abo2023}, we have constructed a
fundamentally new setup, which enabled us to experimentally
implement an Asb\'oth \textit{et al.} entanglement
potential~\cite{Asboth2005} (specifically we studied the
concurrence potential) and the two recently proposed potentials for other
types of quantum correlations for single-qubit states i.e., a steering potential in a three-measurement scenario and a Bell nonlocality potential \cite{Miranowicz2023}. 
This allowed us to observe the hierarchy of NC potential of single-qubit states. 
We have demonstrated several counter-intuitive regimes of this hierarchy. Most notably, upon the interpolation between two states, one of the three tested NC potentials might have opposite trend (derivative sign) than the remaining two potentials. Moreover, we have experimentally observed that states of even a relatively high concurrence potential can be nonsteerable, while other experimental states of a considerably lower concurrence were found both steerable and Bell nonlocal. 
\blue{Various quantum correlations have direct impact on the security of quantum key distribution \cite{Vazirani2014,Branciard2012}. Methods capable of unraveling counterintuitive relations between these types of quantum correlations can lead to better understanding and further research in this direction.}We believe that this research sheds some light on the intricacies of single-qubit nonclassicality and contributes to its better understanding in general.

\section*{Acknowledgments}
J.K  gratefully acknowledges the support from the project IGA\_PrF\_2023\_005 of Palacky University.
A.M. is supported by the Polish National Science Centre (NCN)
under the Maestro Grant No. DEC-2019/34/A/ST2/00081.
We thank Cesnet for data management services.

\section*{Disclosures}

The authors declare no conflicts of interest.

\section*{Data availability} Dataset underlying the results presented in this paper is available in Ref. \cite{DigitalSupplement}.

\section*{Supplemental document}
See digital Dataset (Ref. \cite{DigitalSupplement}) for supporting content.

\section*{Appendix A - Measurement procedure}
In this Appendix, we describe the applied measurement procedure and the output-state reconstruction method in a more detail. 

Instead of encoding qubits into the vacuum ($\ket{\text{vac}}$) and the Fock ($\ket{1}$) state, we use the vertical ($\ket{V}$) and horizontal ($\ket{H}$) polarizations, respectively.
One can think of the vertical polarization as a sort of place-holder or a witness for a vacuum in the horizontal mode.

Considering a pure input qubit state $\ket{\psi_ \text{in}} = \sqrt{1-p}\ket{\text{vac}} + \sqrt{p}\ket{1}$ represented in our experiment by $\ket{\psi_ \text{in}} = \sqrt{1-p}\ket{V}_1 + \sqrt{p}\ket{H}_1$, that is combined on the polarization-dependent beam splitter (PDBS) with the vacuum $\ket{\text{vac}}$, represented by $\ket{V}_2$, then the output state is given by
\begin{equation} \label{output_state}
\begin{split}
\ket{\psi_\text{out}} =
&\frac{1}{\sqrt{2}}\sqrt{p} r_H\ket{H}_4(\ket{V}_3-\ket{V}_4)\\
- &\frac{1}{\sqrt{2}}\sqrt{p} t_H\ket{H}_3(\ket{V}_3-\ket{V}_4)\\
+&\frac{1}{2} \sqrt{1-p}(\ket{V}_3\ket{V}_3-\ket{V}_4\ket{V}_4),
\end{split}
\end{equation}
where $t_H$ and $r_H = \sqrt{1-t_H^2}$ are the transmission and reflection coefficients for the horizontal polarization, respectively, and the subscripts denote the beam splitter port (1 and 2 for the inputs and 3, 4 for the output ports). For the vertical polarization, the PDBS acts as a balanced beam splitter $t_V = r_V = 1/\sqrt{2}$.	The full transformation given by the setup is described in Appendix C  by Eq. (\ref{full_transformation}).
Each term in Eq. (\ref{output_state}) conceptualy corresponds to one possible scenario that can happen when a state of the form of $\ket{\psi_ \text{in}} = \sqrt{1-p}\ket{\text{vac}} + \sqrt{p}\ket{1}$ overlaps on the beam splitter with the vacuum state:  the photon being in (a) port\textsubscript{4} or (b)  port\textsubscript{3}, or (c) no photon present in either of the ports. It is important to notice, that since the vertical-polarization part is just a place-holder for the vacuum, it is irrelevant in which port it is developed.
 Moreover, it is present symmetrically in both output ports which simplifies the detection procedure by detecting the vertically-polarized photon in just one output port and multiplying the corresponding signal accordingly.

When considering a dephased input qubit state instead of a pure one, the output state is described by a density matrix spanned by $\{\ket{00},\ket{01}, \ket{10}, \ket{11}\}$. Because the other input state is the vaccum, it is impossible to detect two photons, so all the terms, corresponding to the term $\ket{11}$ are zero, allowing us to reduce the density matrix
\begin{equation} \label{rho_out}
\rho_{\text{out}} =
\begin{blockarray}{ccccc}
&\ket{00} & \ket{01} & \ket{10} & \ket{11}\\
\begin{block}{c@{\hspace{.3cm}}(cccc)}
\bra{00} & \rho_{11} & \rho_{12} & \rho_{13} & 0\\
\bra{01} & \rho_{12}^* & \rho_{22} &\rho_{23} &0\\
\bra{10} & \rho_{13}^* &\rho_{23}^* &\rho_{33} &0 \\
\bra{11} & 0 &0 & 0& 0 \\
\end{block}
\end{blockarray}\text{ }
\to
\begin{blockarray}{cccc}
&\ket{00} & \ket{01} & \ket{10} \\
\begin{block}{c@{\hspace{.3cm}}(ccc)}
\bra{00} & \rho_{11} & \rho_{12} & \rho_{13} \\
\bra{01} & \rho_{12}^* & \rho_{22} &\rho_{23}\\
\bra{10} & \rho_{13}^* &\rho_{23}^* &\rho_{33}  \\
\end{block}
\end{blockarray}\text{ }.
\end{equation}
We can divide the density matrix into several parts

\begin{equation}
\label{block_matrix}
\begin{aligned}
\rho_{\text{out}} =
&\begin{array}{ccc}
\begin{pmatrix}
M_A & M_C & M_D  \\
M_C^*  & \multicolumn{2}{c}{\multirow{2}{*}{\huge $M_B$\normalsize}} \\
M_D^* &  & \\
\end{pmatrix},
\end{array}
\end{aligned}
\end{equation}
which are all measured using different settings.

The $M_A$ element corresponds to the last line of Eq. (\ref{output_state}) (corresponding to case (c) with no photon) describing a superposition of bunched verticaly polarized photons in both output ports. In this case, we restrain ourself just to the term in port\textsubscript{3}. HWP\textsubscript{3} is rotated by 22.5°, which ensures that photons are separated on PBS\textsubscript{3} with 50\% probability. The signal is measured by the coincidence detection on the detectors Det\textsubscript{A} and Det\textsubscript{B}, while the shutter in port\textsubscript{4} is closed. To correct for the signal that was not detected, the coincidence counts are multiplied by the factor of four (by two for the splitting 50\% probability and by additional two for the photons bunched in port\textsubscript{4}).
Note that there cannot be detected any unwanted signals in this configuration. We limit ourself just to photons in port\textsubscript{3} by closing the shutter, which makes it impossible to detect the state corresponding to the first term in Eq. (\ref{output_state}). The second term in Eq. (\ref{output_state}) contains a component $\ket{H}_3\ket{V}_3$ corresponding to two photons in port\textsubscript{3}. However, that cannot yield any coincidence counts, because the two photons bunch on PBS\textsubscript{3} due to the rotation of HWP\textsubscript{3} and the Hong-Ou-Mandel effect.

The $M_B$ submatrix corresponds to the first two terms in Eq. (\ref{output_state}) and the coherence between them. From these two terms, we focus on the cases, where the two photons are in different ports,

\begin{equation}
\ket{\psi_{B}} = \frac{1}{\sqrt{2}}(\sqrt{p}r_H\ket{H}_4\ket{V}_3 -
\sqrt{p}t_H\ket{H}_3\ket{V}_4).
\end{equation}

We performed a full two-qubit tomography \cite{Jezek2003} using the detectors Det\textsubscript{A} and Det\textsubscript{C} and estimated a matrix $\rho_B$ using the maximum likelihood  estimation \cite{Hradil1997} as

\begin{equation}
\label{matrix_II}
\begin{aligned}
\rho_{B} =
&\begin{array}{cccc}
\begin{pmatrix}
0 & 0 & 0 & 0 \\
0  & \multicolumn{2}{c}{\multirow{2}{*}{\huge$ \tilde{M}_B$\normalsize}} &0\\
0 &  & &0\\
0 & 0 & 0 & 0\\
\end{pmatrix}
\end{array}
\end{aligned},
\end{equation}
where the middle submatrix is just a normalized $M_B$ matrix from Eq. (\ref{block_matrix}). The proper renormalization of the $M_B$ matrix is calculated from the two tomography measurements that correspond to the diagonal elements of the $M_B$ matrix and the value measured for the $M_A$ component. It is importatnt to remember that also the coincidences for the $M_B$ matrix are multiplied by a factor two.
Note that due to experimental imperfections, the two-qubit tomography produces non zero elements all across the matrix in Eq. (\ref{matrix_II}), but the values outside of $\tilde{M}_B$ are typically below 0.05 which justify their neglection.

Note, that the measurement of $M_A$ and $M_B$ are done using different sets of detectors (det$_A$ and det$_B$ for $M_A$, while det$_A$ and det$_C$ for $M_B$). Due to various factors (e.g. the detector efficiency, fiber coupling efficiency and also the loss of a signal on the fiber beam splitter), the overall detection efficiencies of both sets of detectors are different. 
To obtain the correct results, it is necessary to establish them (or at least their ratio),which was done 
 by setting the wave-plates, so that maximal signal is on the desireds set of detector and measuring the signal intensity (coincidence counts).

The coherence terms $M_C$ and $M_D$ are both measured similarly based on visibility in the interference pattern formed by the second polarizing beam splitter in the PDBS and the fiber beam splitter (FBS). 

Let us assume a density matrix in an orthonormal basis $\ket{\psi_1}$ and $\ket{\psi_2}$ to be
\begin{equation}
\begin{blockarray}{ccc}
&\ket{\psi_1} & \ket{\psi_2} \\
\begin{block}{c@{\hspace{.3cm}}(cc)}
\bra{\psi_1} & a & b  \\
\bra{\psi_2} & b* & c \\
\end{block}
\end{blockarray}\text{ },
\end{equation}
that might be unnormalised and can be just a submatrix of a bigger matrix. 
By implementing measurements along $1/\sqrt{2}(\ket{\psi_1}+e^{i\phi}\ket{\psi_2})$ and plotting the signal intensity as a function of $\phi$, one obtains a sinusoidal relation and can calculate the visibility $v = (I_{\text{max}}-I_{\text{min}})/(I_{\text{max}}+I_{\text{min}})$, where $I_{\text{max}}$ ($I_{\text{min}}$) is the maximal (minimal) value of intensity (i.e., the number of coincidence counts). The visibility defined this way is related directly to the absolute value of the off-diagonal element $|b|= v/[2(a+c)]$ reducing into $|b|= v/2$, when the matrix is normalized. Also the phase of $b$ could be reconstructed this way from the position of the maximum.
In our setup, the phase is not stable, so it is only possible, to search for the maximum and minimum, but not to reconstruct the whole sinusoidal pattern, allowing only to measure the absolute value of $b$. For a general matrix, that would not be enough, but in our case, it contains a sufficient information due to the fact that the
 density matrix, as given by Eq.~(\ref{rhorqpx}), can by transformed into a density matrix with only real positive numbers including off-diagonal terms by using just local rotations, which do not affect any measures of quantum correlations. Thus without loss of generality, we choose the phases of the off-diagonal elements to be zero.

For the measurement of the $M_D$ term, the same configuration in port\textsubscript{3} can be set as when measuring the $M_A$ term (i.e., HWP\textsubscript{3} rotated by 22.5°) and projecting half of the term $\ket{V}_3\ket{V}_3$ onto the detectors det\textsubscript{A} and det\textsubscript{B}.
The wave-plates in port\textsubscript{4} are set to zero and the shutter is open which along with the configuration in port\textsubscript{3} projects half of the term $\ket{H}_4\ket{V}_3$ onto the same set of detectors. The FBS in front of det\textsubscript{B} causes these two terms to interfere and the phase between the paths is tuned using a piezo actuator PA\textsubscript{3}, which permits to sweep several times through the entire range of phase shifts. A series of 50 coincidence counts (due to the setup stability accumulated for 5~s each) is registered during this sweep, and the minimal and maximal values from this set are used for the visibility calculation.

The term $M_C$ is detected in the same way, only with symmetrically swapped rotation settings in the output ports (all the wave-plates rotated by 0° except for HWP\textsubscript{4}, which is rotated by 22.5°).

\section*{Appendix B - Estimating the output density matrices}

A complex maximum likelihood-based procedure is used to estimate a physically valid \cite{James2001} output state density matrix $\rho_\mathrm{out}$ correspoging to  Eq.~(\ref{block_matrix}). This procedure takes into account that, due to experimental conditions, the uncertainty in estimating the terms $M_A$ and $M_B$ is about one order of magnitude smaller that the uncertainty of the terms $M_C$ and $M_D$ (requiring visibility measurements). We therefore first estimate the terms $M_A$ and $M_B$ and subsequently keep them fixed while identifying the proper values of the terms $M_C$ and $M_D$.

As already explained in the previous subsection, the submatrix $M_B$ is obtained from the maximum likelihood procedure based on respective polarization projections across output ports. Meanwhile the value of the $M_A$ term is obtained by a direct measurement. Note that its value arises from one measurement only, so in this case, a direct calculation is equivalent to a maximum likelihood procedure. Accumulating the signal for 50~s for each projection, the values of $M_A$ and $M_B$ are the most trusted.

The terms $M_C$ and $M_D$ come from a series of shorter (5s-long, as explained in the preceding subsection) measurements and, as a result, the calculated visibility is accompanied by a significantly higher uncertainty in comparison to the terms $M_A$ and $M_B$. The most likely values of $M_C$ and $M_D$ are estimated from the visibility measurements conditioning on the physicality of the entire output state density matrix $\rho_\mathrm{out}$ with the terms $M_A$ and $M_B$ already fixed.  As the physicality condition for this maximum likelihood estimation, we use Sylvester's criterion for testing the semidefiniteness of a Hermitian matrix \cite{Horn2012}. This condition states that if every principal minor of a matrix is nonnegative, then the matrix is positive semidefinite.
We obtain the following 7 conditions depending on the dimensionality of the the submatrix:
\begin{align*}
(a)&\; \rho_{jj} \ge 0 \;\;\; j=\{1,2,3\}; \\
(b)&\; \rho_{jj}\rho_{kk} -\rho_{jk}^2 \ge 0  \;\;\; j,k=\{1,2,3\}; \;j\ne k;\\
(c)& \; \rho_{11}\rho_{22}\rho_{33}+2\rho_{12}\rho_{23}\rho_{13}
-\rho_{11}\rho_{23}^2-\rho_{22}\rho_{13}^2-\rho_{33}\rho_{12}^2 \ge 0,
\end{align*} 
while assuming the off-diagnoal elements, to be real, as explained above in the main text.
The conditions (a) for the diagonal elements of $\rho$ are already satisfied, as is the condition (b) for $j=2$ and $k=3$ (i.e., the semidefiniteness of the matrix $M_B$). To satisfy the remaining conditions in (b), for the off-diagonal terms, first we replace $\rho_{12}$ with $\rho_{12}'$ given by
\begin{equation}
\rho_{12}' =  \text{min}(\rho_{12},\sqrt{\rho_{11}\rho_{22}}),
\end{equation}
and we apply the same procedure with $\rho_{13}$. This way, we can satisfy the condition (b) . To fulfill the last condition (c), we estimate a parameter
\begin{equation}
c^2 =  -\frac{\rho_{11}\rho_{22}\rho_{33}-\rho_{11}\rho_{23}^2}{2\rho_{12}\rho_{23}\rho_{13}-\rho_{22}\rho_{13}^2-\rho_{33}\rho_{12}^2}
\end{equation}
used as a physicality correction of both elements $\rho_{12}'$ and $\rho_{13}'$:  if the condition (c) is not satisfied, then both $\rho_{12}'$ and $\rho_{13}'$ are multiplied by $c$.

\section*{Appendix C - Full setup transformation of the input states}

The following equations describe the transformation by individual components of the setup of a pure input state on the unbalanced and nondecohering BS
\begin{equation} \label{full_transformation}
\begin{split}
&\ket{V}_1\ket{V}_2 \\
\overset{\text{HWP\textsubscript{1}}}{\longrightarrow}
&(\sqrt{p}\ket{H}_1+\sqrt{1-p}\ket{V}_1)\ket{V}_2\\ 
\overset{\text{HWP\textsubscript{2}}}{\longrightarrow}
&(\sqrt{p}\ket{V}_1+\sqrt{1-p}\ket{H}_1)\ket{V}_2\\  
\overset{\text{PBS\textsubscript{1}}}{\longrightarrow}
&-\sqrt{p}\ket{V}_H\ket{V}_V - \sqrt{1-p}\ket{H}_V\ket{V}_V\\ 
\overset{\text{HWP\textsubscript{H,V}}}{\longrightarrow}
&\sqrt{p}/\sqrt{2}(r_H\ket{H}_H+t_H\ket{V}_H)(\ket{H}_V-\ket{V}_V) + 
\sqrt{1-p}/\sqrt{2}(\ket{H}_V\ket{H}_V - \ket{V}_V\ket{V}_V)\\
\overset{\text{PBS\textsubscript{2}}}{\longrightarrow}
&\sqrt{p}/\sqrt{2}(r_H\ket{H}_4-t_H\ket{V}_3)(\ket{H}_3-\ket{V}_4) + 
\sqrt{1-p}/\sqrt{2}(\ket{H}_3\ket{H}_3 - \ket{V}_4\ket{V}_4)\\
\overset{\text{HWP\textsubscript{3}}}{\longrightarrow}
&\sqrt{p}/\sqrt{2}(r_H\ket{H}_4-t_H\ket{H}_3)(\ket{V}_3-\ket{V}_4) + 
\sqrt{1-p}/\sqrt{2}(\ket{V}_3\ket{V}_3 - \ket{V}_4\ket{V}_4);
\end{split}
\end{equation}
which is obtained by considering transformation matrix of PBS,
\begin{equation}
\begin{pmatrix}
a_{H_{1,\text{out}}}\\
a_{V_{1,\text{out}}}\\
a_{H_{2,\text{out}}}\\
a_{V_{2,\text{out}}}
\end{pmatrix}
=
\begin{pmatrix}
1& 0 & 0 & 0\\
0& 0 & 0 & 1\\
0& 0 & 1 & 0\\
0& -1 & 0 & 0
\end{pmatrix}
\begin{pmatrix}
a_{H_{1,\text{in}}}\\
a_{V_{1,\text{in}}}\\
a_{H_{2,\text{in}}}\\
a_{V_{2,\text{in}}}
\end{pmatrix},
\end{equation}
and HWP rotated by $\theta$,
\begin{equation}
\begin{pmatrix}
a_{H_{\text{out}}}\\
a_{V_{\text{out}}}\\
\end{pmatrix}
=
\begin{pmatrix}
\text{cos}(2\theta)& \text{sin}(2\theta)  \\
\text{sin}(2\theta)& -\text{cos}(2\theta)\\
\end{pmatrix}
\begin{pmatrix}
a_{H_{\text{in}}}\\
a_{V_{\text{in}}}\\
\end{pmatrix}.
\end{equation}

\hfill \break

\bibliography{Hierarchy_potentials}

\end{document}